# Doppler effect in vacuum and medium in X-ray range

A. Shchagin[1,2,*], G. Kube[1], A. Potylitsyn[3], S. Strokov[1]

[1]Deutsches Elektronen-Synchrotron DESY, Notkestrasse 85, 22607 Hamburg, Germany
[2]Kharkov Institute of Physics and Technology, Academicheskaya 1, Kharkiv 61108, Ukraine
[3]Institute of Applied Problems of Physics, 25, Hr. Nersisyan Str., 0014, Yerevan, Republic of Armenia

*Corresponding author, e-mail: alexander.shchagin@desy.de

**Abstract**
This paper discusses the Doppler effect for radiation emitted by a relativistic oscillator in vacuum described geometrically as an ellipsoid in momentum space. Spectral and angular properties of the Doppler X-ray radiation emitted by a relativistic oscillator moving through a homogeneous medium are considered. Additionally, we show that radiation from an elementary relativistic charged particle interacting with periodic medium can be interpreted as a manifestation of the Doppler effect. Examples include parametric X-ray radiation, coherent bremsstrahlung, undulator radiation in crystalline or vacuum undulators, transition radiation from stacked foils, and Smith-Purcell radiation.

**Key words:** Doppler effect, X-ray radiation, homogeneous and periodical media

**1. Introduction**

The Doppler effect is usually studied for oscillators moving in a vacuum or a homogeneous medium [1,2]. In this work, we discuss the Doppler formula for relativistic oscillator moving in vacuum and show that it forms an ellipsoid in momentum space. Besides, we derive the analytical expressions for description of the spectral and angular properties of X-ray radiation emitted by a relativistic oscillator moving through a homogeneous medium.

In his monograph [3], Ter-Mikaelian derived the equation for the frequency of parametric X-ray radiation (PXR) produced by the motion of a relativistic elementary charged particle in a crystal. This derivation was based on energy and momentum conversation laws. Later, in [4] he noted that the same equation for PXR frequency could also be derived from the Doppler formula. This work further explores the Doppler effect as a unifying framework for various types of X-ray radiation arising from the interaction of relativistic charged particles with periodic structures. Examples include parametric X-ray radiation, coherent bremsstrahlung, undulator radiation (in crystalline or vacuum undulators), transition radiation from stacked foils, and Smith-Purcell radiation.

**2. Doppler effect in a vacuum as an ellipsoid in the momentum space**

The radiation frequency of an oscillator with angular eigenfrequency $\omega_0$ which moves rectilinearly and uniformly with velocity $V$ in a vacuum, is described by the Doppler formula [1]. For an observer in the laboratory frame, the radiation appears at a single frequency $\omega_V$

$$\omega_V = \frac{\omega_0 \gamma^{-1}}{1 - \frac{V}{c}\cos\theta}, \qquad (1)$$

where $\gamma = \frac{1}{\sqrt{1-\left(\frac{V}{c}\right)^2}}$ is the relativistic Lorentz factor, $c$ is the speed of light, and $\theta$ is the observation angle relative to the particle velocity vector $\vec{V}$. Note that the formula for an ellipsoid in polar coordinates has a very similar shape [5]

$$\rho = \frac{d}{1 - \xi \cos\theta}, \qquad (2)$$

where $\xi$ is the eccentricity of the ellipsoid, $\rho$ is the distance from the ellipsoid's focal point to its surface, $\theta$ is the polar angle with respect to the long ellipsoid diameter, $d = \frac{b^2}{a}$.

Rewrite the Doppler formula (1) as a function of the relativistic Lorentz factor

$$\left|\vec{k}\right| = \frac{\hbar \omega_V}{c} = \frac{\hbar \omega_0 \gamma^{-1}}{c\left(1 - \sqrt{1 - \gamma^{-2}}\cos\theta\right)}, \qquad (3)$$

where $\vec{k}$ is the wave vector of the emitted radiation, $\hbar$ is the reduced Plank constant. Similarly, rewrite the ellipsoid formula (2) as a function of the long $a$ and short $b$ semi-diameters of the ellipsoid

$$\rho = \frac{b^2/a}{1 - \sqrt{1-\left(\frac{b}{a}\right)^2}\cos\theta}. \qquad (4)$$

Comparing the Doppler and ellipsoid formulas, one can see that they have the same structures. By comparing their parameters, the following correspondences can be seen: **i.** the wave vector module $\left|\vec{k}\right| = \rho$; **ii.** the particle velocity in units of the speed of light is equivalent to the ellipsoid eccentricity $\frac{V}{c} = \xi$; **iii.** the relativistic Lorentz factor is equal to the ratio of the ellipsoid's long to short diameters $\gamma = \frac{a}{b}$; **iv.** the coefficient $\frac{\hbar \omega_0}{c}$ is equal to the ellipsoid's short semi-diameter $\frac{\hbar \omega_0}{c} = b$. Thus, the Doppler formulas in a vacuum (1, 3) represent an ellipsoid in momentum space as shown in Fig. 1.

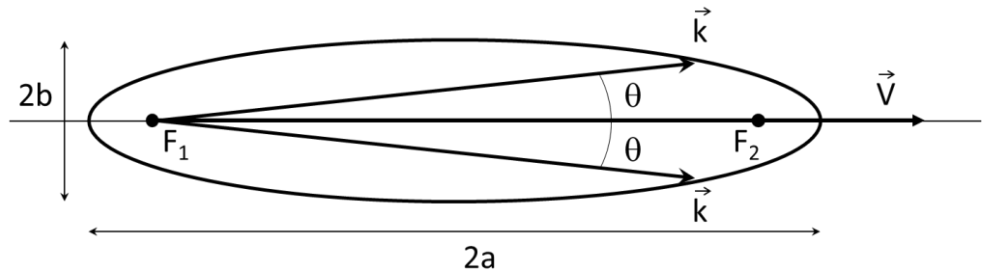

Fig. 1. Cross-section of the Doppler ellipsoid in vacuum in momentum space. The origin of the polar coordinates is located at the left focal point $F_1$. The relativistic oscillator moves with velocity

$\vec{V}$. The ellipsoid defined by its focal points $F_1$ and $F_2$ is characterized by its long $2a$ and short $2b$ diameters. The relativistic Lorentz factor of the oscillator is given by the relation $\gamma = \dfrac{a}{b}$.

The Doppler ellipsoid is strongly elongated along the velocity vector $\vec{V}$ at $\gamma \gg 1$. In vacuum, radiation occurs at only one frequency (Eq. (1)) regardless of the oscillator's angular frequency $\omega_0$, velocity $V$, or Lorentz factor $\gamma$. The maximum frequency is emitted in the forward direction at $\theta = 0$:

$$\omega_V(\theta = 0) = \gamma \omega_0 \left(1 + \frac{V}{c}\right). \tag{5}$$

## 3. The Doppler effect in a medium in the X-ray range

Let us consider the radiation frequency of the same relativistic oscillator with angular eigenfrequency $\omega_0$ moving through a homogeneous medium. If the particle moves in a medium with permittivity $\varepsilon$, equation (1) is modified as [1]

$$\omega = \frac{\omega_0 \gamma^{-1}}{1 - \dfrac{V\sqrt{\varepsilon}}{c}\cos\theta}. \tag{6}$$

In the following discussion, we focus on radiation in the X-ray and gamma-ray ranges, where

$$\varepsilon = 1 - \left(\frac{\omega_p}{\omega}\right)^2, \tag{7}$$

$\omega_p$ is the plasma frequency in the medium, and emitted frequency $\omega$ lies outside of the resonant atomic or nuclear frequencies in the medium. Additionally, the emitted frequency must satisfy

$$\omega > \omega_p, \tag{8}$$

because radiation with $\omega < \omega_p$ is absorbed within the medium. It is worth noting that the denominator in Eq. (6) remains positive in the X-ray range. For a discussion on the sign of this denominator in non-X-ray ranges, refer to [1].

The Cherenkov radiation is not possible in this case because the phase velocity of light $\dfrac{c}{\sqrt{\varepsilon}}$ in the X-ray range exceeds the speed of light in vacuum $c$, as well as the velocity of oscillator $V$. Substituting Eq. (7) into Eq. (6), we obtain a quadratic equation for the frequency of the radiation emitted by the oscillator:

$$\omega^2\left[1 - \left(\frac{V\cos\theta}{c}\right)^2\right] - \omega 2\omega_0 \gamma^{-1} + \omega_0^2 \gamma^{-2} + \omega_p^2\left(\frac{V\cos\theta}{c}\right)^2 = 0. \tag{9}$$

Eq. (9) has two solutions

$$\omega_{+/-} = \frac{\omega_0 \gamma^{-1}\left[1 \pm \dfrac{V\cos\theta}{c}\sqrt{1 - \left(\dfrac{\omega_p \gamma}{\omega_0}\right)^2\left[1 - \left(\dfrac{V\cos\theta}{c}\right)^2\right]}\right]}{\left[1 - \left(\dfrac{V\cos\theta}{c}\right)^2\right]} = \omega_V \frac{1 \pm \dfrac{V\cos\theta}{c}\sqrt{1 - \left(\dfrac{\omega_p \gamma}{\omega_0}\right)^2\left[1 - \left(\dfrac{V\cos\theta}{c}\right)^2\right]}}{1 + \dfrac{V\cos\theta}{c}}.$$

(10)

One can see that the oscillator in a medium can emit two frequencies in the X-ray range instead of only one frequency in a vacuum. These are the high-frequency solution or branch $\omega_+$ and the low-frequency solution or branch $\omega_-$. Radiation in the low-frequency branch is only possible if $\omega_-$ exceeds the plasma frequency $\omega_p$, as specified by condition (8).

Ginzburg, in [1], defined the anomalous Doppler effect as radiation emitted by an oscillator moving at a velocity greater than the phase velocity of light. However, in the X-ray range, this phenomenon is not possible because the phase velocity of light always exceeds the speed of light in vacuum, as noted above.

For the two solutions derived in equation (10), we refer to them as the high-frequency and low-frequency branches of radiation. The physical meaning of the high-frequency branch can be understood using the Huygens principle, where it arises from the constructive interference of waves emitted by the oscillator. The physical interpretation of radiation in the low-frequency branch (the solution with the negative sign in equation (10)) remains an open question. In the following, we will focus on radiation emitted in the forward hemisphere.

Radiation at the frequencies described by Eq. (10) is possible if the radicand in Eq. (10) is non-negative:

$$\left(\frac{\omega_p \gamma}{\omega_0}\right)^2 \left(1 - \left(\frac{V \cos\theta}{c}\right)^2\right) \leq 1. \tag{11}$$

From inequality (11), the range of angles where radiation is possible can be determined:

$$|\sin\theta| \leq |\sin\theta_{max}| = \frac{c}{V\gamma}\sqrt{\left(\frac{\omega_0}{\omega_p}\right)^2 - 1}, \tag{12}$$

where

$$|\theta_{max}| = \arcsin\left[\frac{c}{V\gamma}\sqrt{\left(\frac{\omega_0}{\omega_p}\right)^2 - 1}\right] \tag{13}$$

is the maximum angle at which radiation is possible. An additional condition for the existence of radiation, beyond condition (8) follows from Eq. (12):

$$\omega_0 > \omega_p. \tag{14}$$

Condition (12) imposes restrictions on the spectral and angular distributions of the emitted radiation. Two frequencies from Eq. (10) are emitted within the observation angle range $0 \leq \theta < \theta_{max}$. However, only one frequency

$$\omega = \gamma \frac{\omega_p^2}{\omega_0} \tag{15}$$

is emitted when $\theta = \theta_{max}$ (when the radicand in Eq. (10) is zero). No radiation is emitted at $\theta > \theta_{max}$. The frequencies of radiation emitted in the forward direction at $\theta = 0$ are:

$$\omega_{+/-}(\theta = 0) = \omega_0 \gamma \left[1 \pm \sqrt{(1 - \gamma^{-2})\left(1 - \left(\frac{\omega_p}{\omega_0}\right)^2\right)}\right]. \tag{16}$$

In the case when $\gamma \gg 1$ and $\omega_0 \gg \omega_p$, the two frequencies (16) emitted in the forward direction at $\theta = 0$ are:

$$\omega_+ (\theta = 0) \approx \gamma \omega_0 \left\{ 2 - \frac{1}{2} \left[ \gamma^{-2} + \left( \frac{\omega_p}{\omega_0} \right)^2 \right] \right\}, \tag{17}$$

$$\omega_- (\theta = 0) \approx \frac{\gamma \omega_0}{2} \left[ \gamma^{-2} + \left( \frac{\omega_p}{\omega_0} \right)^2 \right]. \tag{18}$$

The low-frequency branch of radiation (Eqs. (10), (16), and (18)) can exist in the X-ray range if it satisfies condition (8). In the ultra-relativistic case, at $\gamma \gg 1$, $\gamma \gg \frac{\omega_0}{\omega_p}$, $\omega_0 \gg \omega_p$, the condition (12) in small-angle approximation becomes

$$|\theta| < \frac{\omega_0}{\gamma \omega_p}. \tag{19}$$

Thus, the ultra-relativistic oscillator in a medium can emit X-ray radiation in the forward hemisphere within the angles satisfying equation (19). This behavior is similar to bremsstrahlung radiation emitted by a relativistic charged particle in a medium within angles $|\theta|$ of the order $\sim \gamma^{-1}$.

The ideal vacuum Doppler ellipsoid (described by Eq. (1) and shown in Fig. 1) is distorted in a medium because the "eccentricity" $\frac{V\sqrt{\varepsilon}}{c}$ in Eq. (6) depends on the frequency $\omega$.

## 4. Discussion

Let us discuss the Doppler effect for an elementary charged particle moving in a periodic medium. Usually, an oscillator is considered as a particle without charge with eigenfrequency $\omega_0$ [1, 2]. Ter-Mikaelian was the first to highlight that the frequency of PXR emitted in a crystal and transition radiation emitted in a stack of foils by an elementary charged particle can be considered as a result of the Doppler effect [4]. Below the discussion focuses on radiation emitted by a single charged particle moving through a periodic medium. We believe that all types of resonant (coherent) radiation emitted by a fast charged particle in such a medium can be considered as manifestation of the Doppler effect. The fundamental harmonic frequencies in the emitted spectrum are determined by the frequency of disturbances in the particle's trajectory or its electromagnetic field within the medium. In this case, the Doppler formula (6) can be modified by substituting $\omega_0 \gamma^{-1}$ with $\frac{2\pi V}{l}$, where $l$ is the period of the medium along the particle velocity vector in the laboratory frame. As a result, this substitution leads to the Ter-Mikaelian equation for the PXR mechanism (formula (20), shown below). Other radiation mechanisms in periodic media, such as coherent bremsstrahlung, undulator radiation, and Smith-Purcell radiation, can be interpreted in a similar manner. The following examples illustrate this interpretation.

**Parametric X-ray radiation** arises from the interaction of a fast charged particle moving rectilinearly with the periodic electron subsystem of a crystal. This type of radiation is because polarization mechanism. The PXR frequency $\omega_{PXR}$ is described by equation [3,6] which is the Doppler formula

$$\omega_{PXR,CB} = \frac{2\pi V}{l} \frac{1}{1 - \frac{V\sqrt{\varepsilon}}{c} \cos\theta}, \tag{20}$$

where $l$ is the distance between neighboring crystallographic planes along the particle's trajectory. PXR is produced only on families of crystallographic planes with non-zero structure factors [7].

**Coherent bremsstrahlung** (at $\hbar\omega_{CB} \ll E$, where $E$ is the energy of incident particle) arises from the interaction of a fast charged particle moving rectilinearly with the periodic nuclear subsystem of a crystal [3]. This type of radiation arises because of bremsstrahlung mechanism. The frequency of coherent bremsstrahlung $\omega_{CB}$ is described by the same Eq. (20) for photon emission angles of the order inverse Lorentz factor [8].

**Undulator radiation from crystalline undulator, based on the channeling effect,** is produced by charged particles such as electrons or positrons moving in a channeling regime in a periodically bent crystal [9, 10]. The mechanism of this radiation is the undulator radiation. The fundamental frequency of radiation $\omega_{UR}$ for a particle undergoing sinusoidal motion is described by the Doppler law:

$$\omega_{UR} = \frac{2\pi}{l_U} \frac{V_z}{1 - \frac{V_z \sqrt{\varepsilon}}{c} \cos\theta}, \qquad (21)$$

where $l_U$ is the undulator period, $V_z$ is the average longitudinal velocity of the particle. Some higher harmonics can be produced if the particle's motion deviates from a purely sinusoidal trajectory. Also, formula (21) with $\varepsilon = 1$ describes the frequency of radiation for a classic magnetic undulator when particles move in a vacuum. Equation (21) is commonly used in the small-angle approximation to describe undulator radiation emitted at small observation angles.

**Transition radiation from a stack of foils** is produced by charged particles crossing a periodic structure consisting of layers of materials with different permittivities [11, 12]. This type of radiation arises due to polarization mechanism. The fundamental frequency of transition radiation $\omega_{TR}$ is described by the same Doppler equation:

$$\omega_{TR} = \frac{2\pi V}{l_{ST}} \frac{1}{1 - \frac{V \sqrt{\varepsilon_{eff}}}{c} \cos\theta}, \qquad (22)$$

where $l_{ST}$ is the period of the structure, and $\varepsilon_{eff}$ is the effective dielectric constant of the structure [11]. Equation (22) is usually used in the small-angle approximation to describe transition radiation emitted at small observation angles.

**Smith-Purcell radiation (SPR) in a medium** arises when a fast charged particle moves in a vacuum near a periodical structure (grating) [13]. This type of radiation arises due to polarization mechanism. The frequency of the fundamental harmonics $\omega_{SP}$ propagating in the grating bulk is described by the same Doppler equation:

$$\omega_{SP} = \frac{2\pi V}{l_{SP}} \frac{1}{1 - \frac{V \sqrt{\varepsilon}}{c} \cos\theta}, \qquad (23)$$

where $l_{SP}$ is the grating period, and $\varepsilon$ is the permittivity of the grating material. Equation (23) turns to the conventional SPR equation with $\varepsilon = 1$ when SPR photons propagate in a vacuum.

Thus, the frequencies of different types of radiation produced by charged elementary particles moving through different periodic structures excited by different radiation mechanisms can all be considered as a manifestation of the Doppler effect. In the X-ray range, the permittivity of all media depends on the square of the emitted frequency as described by Eq. (7). This frequency dependence leads to two solutions for the frequencies in equations (20) – (23). However, the physical interpretation of the second solution with the negative sign in the numerator remains unclear.

## 5. Results

In this paper paper, we described the Doppler effect in a vacuum as an ellipsoid in momentum space and analyzed its spectral and angular properties in a homogenous medium within X-ray and gamma-ray ranges. Additionally, we demonstrated that the frequencies of resonant (coherent) radiation emitted by fast charged particles interacting with periodic medium through different mechanisms can be interpreted as manifestations of the Doppler effect.

Results of this research may have practical applications, particularly in studies of radiation emitted by accelerated radioactive nuclei or excited ions, which can emit radiation either in a vacuum or in a target.

## 6. Acknowledgements

This project has received funding through the MSCA4Ukraine project #1233244, which is funded by the European Union. A.V.S. wish to express his gratitude to V.A. Maisheev, S.V. Trofymenko, I.V. Kyryllin, M.V. Bondarenco for useful discussions.

## 7. Declaration of Interest Statement

The authors declare that they have no known competing financial interests or personal relationships that could have appeared to influence the work reported in this paper.